\documentclass[a4paper,english,runningheads]{llncs}
\usepackage[T1]{fontenc}
\usepackage[latin9]{inputenc}
\usepackage{xcolor}
\usepackage{pdfcolmk}
\usepackage{array}
\usepackage{float}
\usepackage{multirow}
\usepackage{amsmath}
\usepackage{amssymb}
\usepackage{graphicx}
\usepackage{setspace}
\usepackage[all]{xy}
\PassOptionsToPackage{normalem}{ulem}
\usepackage{ulem}

\makeatletter

\pdfpageheight\paperheight
\pdfpagewidth\paperwidth

\providecommand{\tabularnewline}{\\}
\providecolor{lyxadded}{rgb}{0,0,1}
\providecolor{lyxdeleted}{rgb}{1,0,0}

\usepackage[all]{xy}

\usepackage{cite}

\newcommand{\ICC}{\mathsf{ICC}}

\usepackage{babel}

\usepackage{babel}

\makeatother

\usepackage{babel}
\begin{document}

\title{Testing Contextuality in Cyclic Psychophysical Systems of High Ranks}

\titlerunning{Cyclic Psychophysical Systems}

\author{Ru Zhang\textsuperscript{1} and Ehtibar N. Dzhafarov\textsuperscript{2}\textsuperscript{}}

\authorrunning{R. Zhang and E. N. Dzhafarov}

\institute{\textsuperscript{1}Purdue University and Indiana University\\
 zhang617@purdue.edu\\
 $\,$\\
 \textsuperscript{2}Purdue University\\
 ehtibar@purdue.edu\\
 $\,$}

\toctitle{Cyclic Psychophysical Systems}

\tocauthor{R. Zhang, E. N. Dzhafarov}
\maketitle
\begin{abstract}
Contextuality-by-Default (CbD) is a mathematical framework for understanding
the role of context in systems with deterministic inputs and random
outputs. A necessary and sufficient condition for contextuality was
derived for cyclic systems with binary outcomes. In quantum physics,
the cyclic systems of ranks $n=$ 5, 4, and 3 are known as systems
of Klyachko-type, EPR-Bell-type, and Leggett-Garg-type, respectively.
In earlier publications, we examined data collected in various behavioral
and social scenarios, from polls of public opinion to our own experiments
with psychophysical matching. No evidence of contextuality was found
in these data sets. However, those studies were confined to cyclic
systems of lower ranks ($n\leq4$). In this paper, contextuality of
higher ranks ($n=6,8$) was tested on our data with psychophysical
matching, and again, no contextuality was found. This may indicate
that many if not all of the seemingly contextual effects observed
in behavioral sciences are merely violations of consistent connectedness
(selectiveness of influences).

\end{abstract}

\section{Introduction}

Consider a system having two external factors (or inputs) $\alpha$
and $\beta$, which can be deterministically manipulated, and two
random outputs $A$ and $B$ that we interpret as responses to, or
measurements of, $\alpha$ and $\beta$, respectively. The system
can belong to any empirical domain, from quantum physics to behavioral
sciences. If manipulating $\beta$ does not change the marginal distribution
of $A$ and manipulating $\alpha$ does not change the marginal distribution
of $B$, we say that the system is consistently connected. Physicists
traditionally test contextuality by assuming consistent connectedness
(referred to as ``no-signaling,'' ``no-disturbance,'' etc.). However,
even in quantum experiments inconsistent connectedness may occur,
e.g., because of context-dependent errors in measurements. In behavioral
sciences inconsistent connectedness is ubiquitous. The Contextuality-by-Default
(CbD) theory allows one to detect and measure contextuality, or to
determine that a system is noncontextual, irrespective of whether
it is consistently connected \cite{DK2014a,DK2014b,DK2014c,DKL2015,DKC2016,KD2015a,KD2015b,KDL2015}.
In quantum physics, many experiments and theoretical considerations
demonstrate the existence of contextual systems \cite{Bell,KS1967,CHSH1969,Fine1982a,Fine1982b,LG1985,KCBS2008},
including in cases when consistent connectedness is violated \cite{KDL2015}.
By contrast, we found no evidence of contextuality in various social
and behavioral data sets, from polls of public opinion to visual illusions
to conjoint choices to word combinations to psychophysical matching
\cite{DKCZJ2016,DZK2015-1}. 

Most of the experimental studies of contextuality, both in quantum
physics and in behavioral and social sciences, have been confined
to cyclic systems \cite{DKL2015,KDL2015,KD2015a}, in which each entity
being measured or responded to enters in two contexts and each context
contains exactly two entities. In this paper we only deal with cyclic
systems of even ranks, those that can be formed using the paradigm
with two experimental factors (or inputs) $\alpha,\beta$ and two
outputs in response to the two factors. A cyclic system of an even
rank $2n\geq4$ can be extracted from a design in which $\alpha$
and $\beta$ vary on $n$ levels each, denoted $\alpha_{1},\alpha_{2},\ldots,\alpha_{n}$
and $\beta_{1},\beta_{2},\ldots,\beta_{n}$. Out of $n^{2}$ possible
treatments one extracts $2n$ pairs, we call contexts, whose elements
form a cycle, e.g., 
\begin{equation}
\begin{array}{ccccc}
\mathrm{Context\,}1 & \mathrm{Context\,}2 & \ldots & \textnormal{Context }\left(2n-1\right) & \textnormal{Context }2n\\
(\alpha_{1},\beta_{1}) & (\beta_{1},\alpha_{2}) & \ldots, & \left(\alpha_{n},\beta_{n}\right) & \left(\beta_{n},\alpha_{1}\right)
\end{array}.\label{eq: even rank cyclic 1}
\end{equation}
This is a cyclic system of rank $2n$. The outputs of the system corresponding
to these $2n$ contexts are $2n$ pairs of random variables 
\begin{equation}
\left(A_{11},B_{11}\right),\left(B_{21},A_{21}\right),\ldots,\left(A_{nn},B_{nn}\right),\left(B_{1n},A_{1n}\right),\label{eq: even rank cyclic 2}
\end{equation}
where $A_{ij}$ is interpreted as a response to (measurement of )
$\alpha_{i}$ in the context $\left(\alpha_{i},\beta_{j}\right)$,
and $B_{ij}$ is interpreted as a response to (measurement of ) $\beta_{j}$
in the same context, where $i,j\in\left\{ 1,\ldots,n\right\} $. It
is assumed in addition that each random output is binary, with values
denoted $-1,+1$. The random variables $A_{ij}$ and $B_{ij}$ (recorded
in the same context) are jointly distributed, so that, e.g., the joint
probability of $A_{ij}=1$ and $B_{ij}=-1$ is well-defined. However,
according to CbD, any two random outputs recorded in different contexts,
such as $A_{ij}$ and $B_{i'j'}$ or $A_{ij}$ and $A_{i'j'}$, with
$\left(i,j\right)\not=\left(i',j'\right)$, are stochastically unrelated,
have no joint distribution \cite{DK2014c,DKC2016,DKCZJ2016,DKL2015,KD2015b,KDL2015}.

In CbD, the system just described is considered noncontextual if and
only if the $2n$ pairs of random variables in (\ref{eq: even rank cyclic 2})
can be coupled (imposed a joint distribution on) so that any two random
variables responding to the same factor point in different contexts
(i.e., $A_{ij}$ and $A_{ij'}$, or $B_{ij}$ and $B_{i'j}$) are
equal to each other with maximal possible probability, given their
individual distributions \cite{DKL2015,KD2015b,KDL2015,DKC2016,DKCZJ2016,DZK2015-1}.
A necessary and sufficient condition for noncontextuality of a cyclic
system (\ref{eq: even rank cyclic 2}) was derived in Refs. \cite{DKL2015,KD2015a}:
\begin{equation}
\begin{array}{r}
\Delta C=\mathrm{s_{1}}\left(\left\langle A_{11}B_{11}\right\rangle ,\left\langle B_{21}A_{21}\right\rangle ,\ldots,\left\langle A_{nn}B_{nn}\right\rangle ,\left\langle B_{1n}A_{1n}\right\rangle \right)\\
\\
-\ICC-\left(2n-2\right)\leq0,
\end{array}\label{delta C}
\end{equation}
where $\left\langle \cdot\right\rangle $ denotes expected value,
$\mathrm{s_{1}}\left(x_{1},\ldots,x_{k}\right)$ is the maximum of
all linear combinations $\pm x_{1}\pm\ldots\pm x_{k}$ with odd numbers
of minuses, and
\begin{equation}
\begin{array}{l}
\ICC=\left|\left\langle A_{11}\right\rangle -\left\langle A_{1n}\right\rangle \right|+\left|\left\langle B_{11}\right\rangle -\left\langle B_{21}\right\rangle \right|\\
\\
+\ldots+\left|\left\langle A_{n(n-1)}\right\rangle -\left\langle A_{nn}\right\rangle \right|+\left|\left\langle B_{nn}\right\rangle -\left\langle B_{1n}\right\rangle \right|.
\end{array}\label{delta C add}
\end{equation}
If a system is consistently connected, $\ICC$ vanishes.

Experimental studies of cyclic systems in quantum physics were confined
to ranks 3, 4, and 5 (see Refs. \cite{DKL2015,KDL2015} for an overview).
In behavioral and social experiments and surveys the ranks of the
cyclic systems explored were 2, 3, and 4 (see Refs. \cite{DKCZJ2016,DZK2015-1}
for an overview). In the present study, we analyze cyclic systems
of ranks 4, 6, and 8.

\section{Experiments}

\begin{singlespace}
The experimental design and procedure were described in detail in
Ref. \cite{DZK2015-1}. Three different psychophysical matching tasks
were used (Figure 1): dot position reproduction task (Experiment 1(a)
and 1(b)), concentric circles reproduction task (Experiment 2(a),
2(b) and 2(c)), and floral shape reproduction task (Experiment 3(a)
and 3(b)). Each of the seven experiments was conducted on three participants. 

\begin{figure}
\begin{centering}
\includegraphics[scale=0.16]{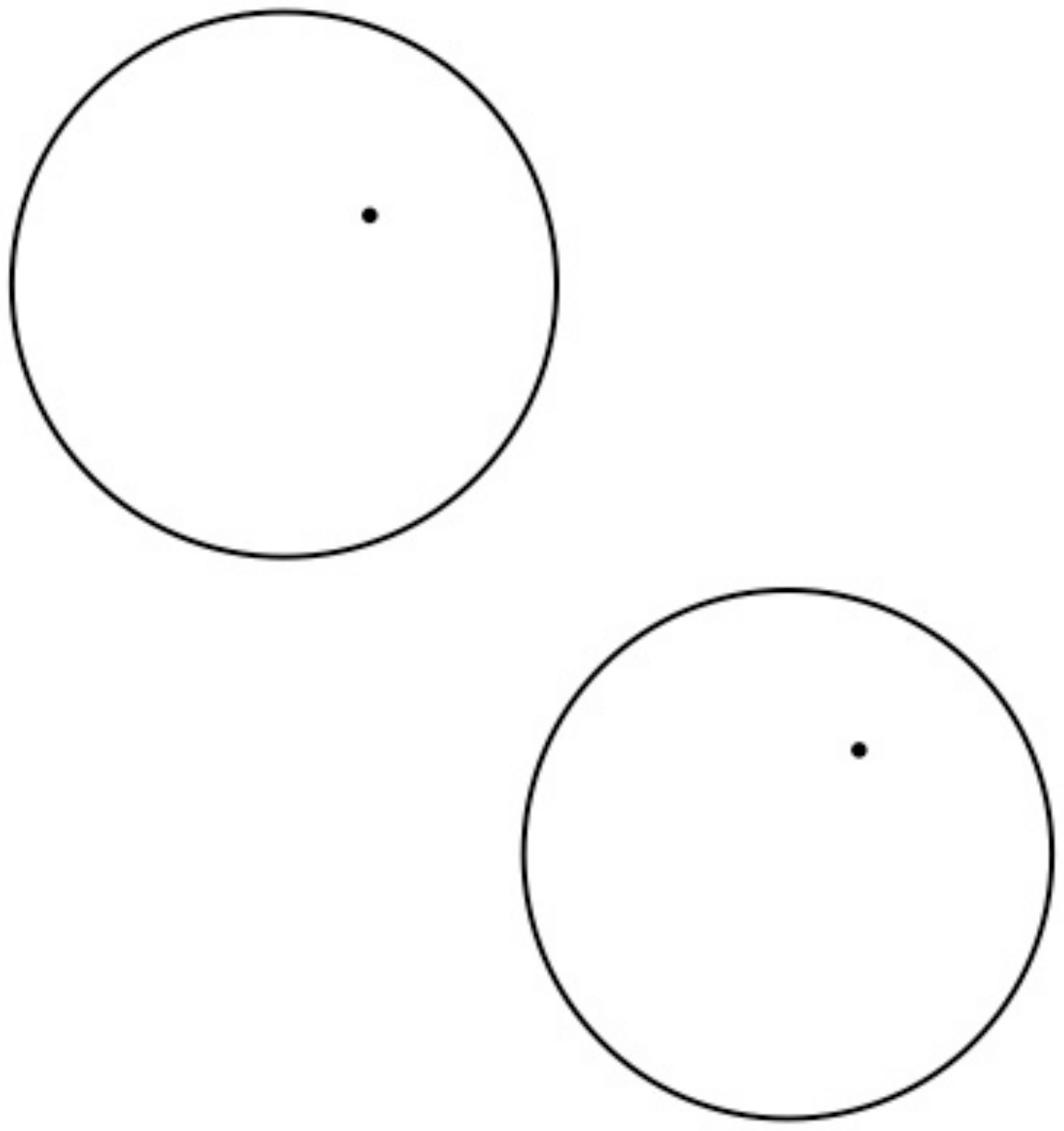}\includegraphics[scale=0.16]{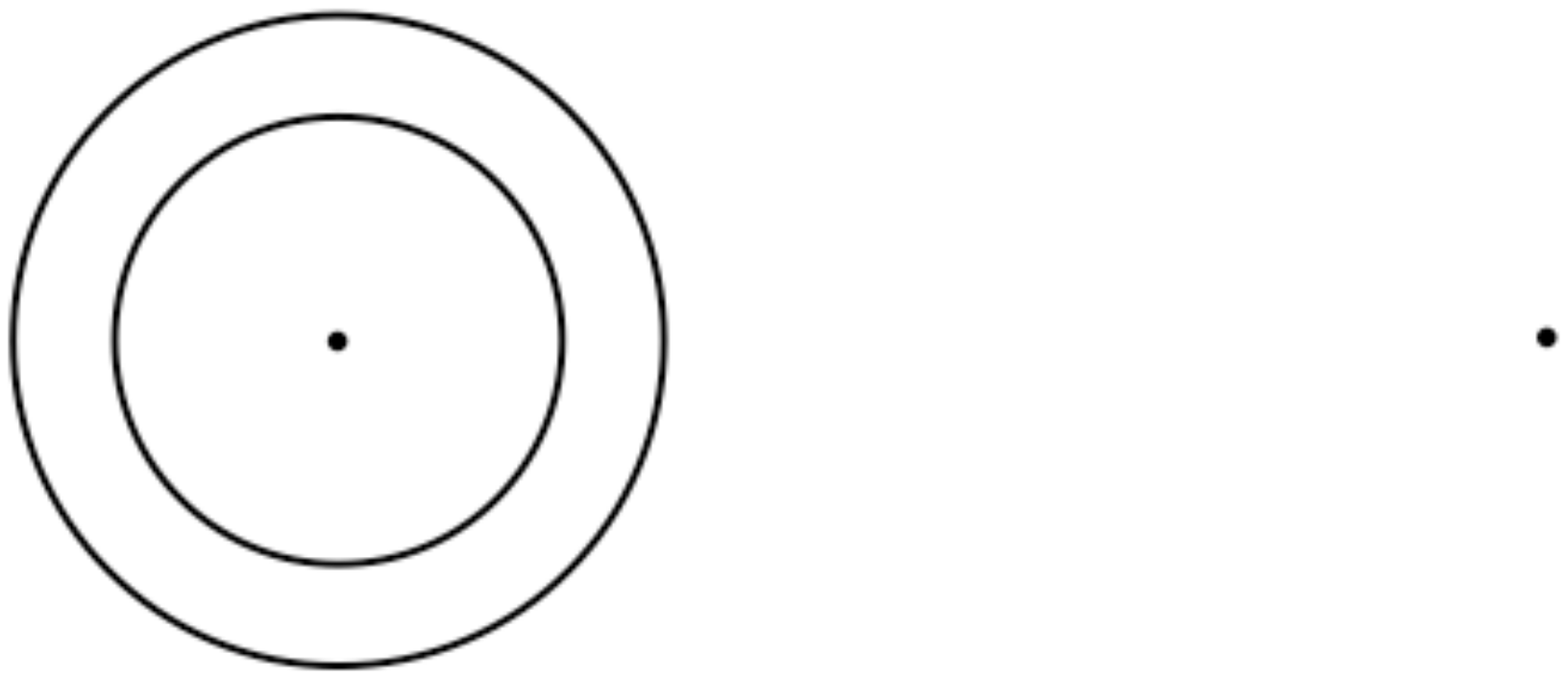}\includegraphics[scale=0.16]{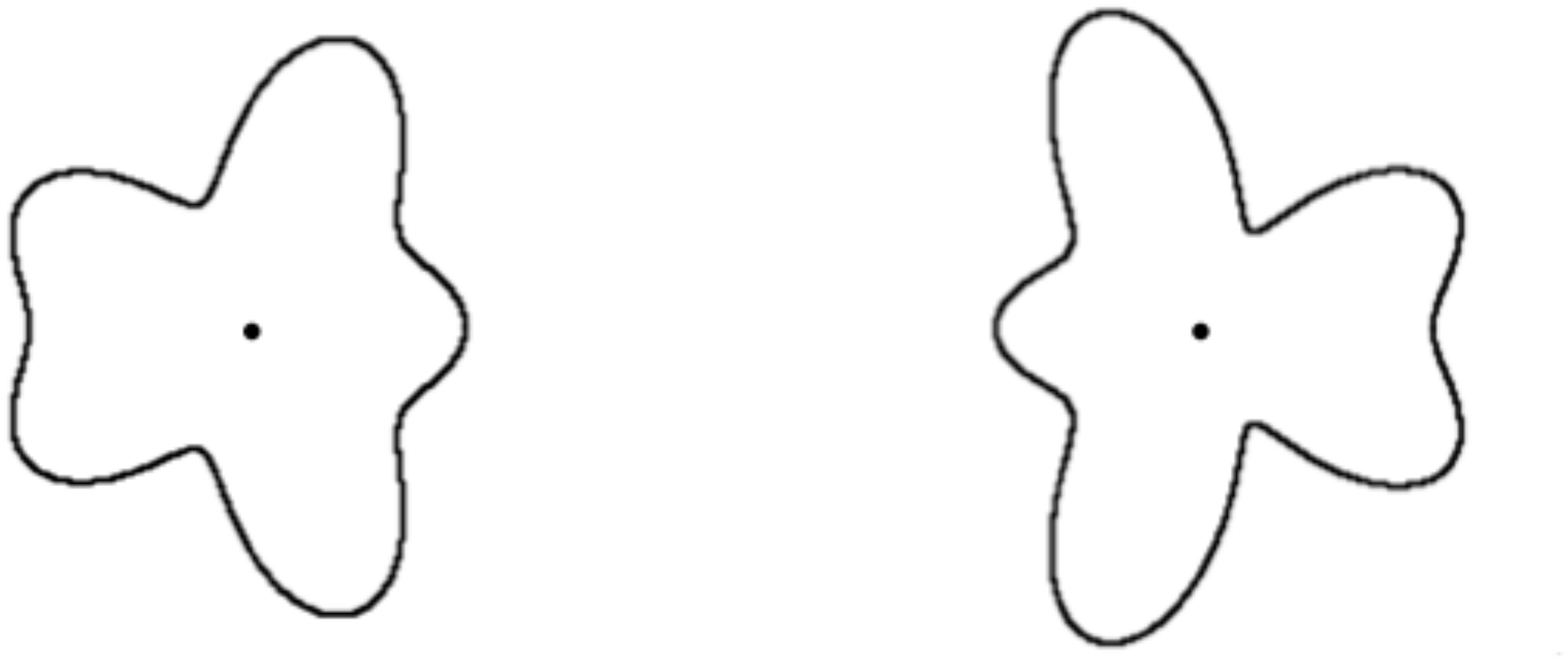} 
\par\end{centering}
\caption{Stimuli used in the (a) dot position reproduction task, (b) concentric
circles reproduction task, and (c) floral shape reproduction task.
\label{fig: Stimuli-used-in}}
\end{figure}

In each experimental trial, the participants were shown two stimuli
on a computer screen, as shown in Fig. \ref{fig: Stimuli-used-in}.
One was a fixed stimulus, the other stimulus was adjustable, by means
of rotating a trackball. The participants were required to change
this stimulus until it appeared to match the position or shape of
the fixed target stimulus. Once a match was achieved, she or he clicked
the button on the trackball to terminate the trial. Each stimulus
was characterized by two parameters. For the target stimulus these
parameters are denoted as $\alpha$ and $\beta$, and their values
in each trial were generated from a pre-defined set of numbers. The
values of the same parameters in the matching stimulus are denoted
$A$ and $B$ (as they randomly vary for given values of $\alpha$
and $\beta$). Table \ref{tab: Factors--and} shows the parameters
used. 
\end{singlespace}

The trials were separated by .5 second intervals. Each experiment
took several days, each of which consisted of about 200 trials with
a break in the middle. Each such session began by a practice series
of 10 trials (which were not used for data analysis). 

\begin{singlespace}
The original data sets for all the experiments are available as Excel
files online (http://dx.doi.org/10.7910/DVN/OJZKKP). Each file corresponds
to one participant in one experiment. 
\end{singlespace}
\begin{singlespace}

\subsection{Participants}
\end{singlespace}

\begin{singlespace}
All the participants were students at Purdue University. The first
author of this paper, labeled as P3, participated in all the experiments.
Participants P1 and P2 participated in Experiments 1(a) and 2(a),
and Participants P4 and P5 in Experiments 1(b), 2(b), 2(c), 3(a),
and 3(b). All participants were about 25 years old and had normal
or corrected to normal vision. 

\begin{table}
\caption{External factors $(\alpha,\beta)$ and random outputs $(A,B)$ for
the three types of tasks.\label{tab: Factors--and}}

\centering{}%
\begin{tabular}{|c|c|c|c|c|}
\hline 
Task & $\alpha$ & $\beta$ & $A$ & $B$\tabularnewline
\hline 
Dot position & Horizontal  & Vertical & Horizontal  & Vertical\tabularnewline
reproduction & coordinate & coordinate & coordinate of & coordinate of\tabularnewline
(rectangular  & of the  & of the & the matching & the matching\tabularnewline
coordinates) & target dot & target dot & dot & dot\tabularnewline
\hline 
Dot position & Radial & Angular & Radial & Angular\tabularnewline
reproduction & coordinate & coordinate & coordinate of & coordinate of \tabularnewline
(polar  & of the & of the  & the matching & the matching\tabularnewline
coordinates) & target dot & target dot & dot & dot\tabularnewline
\hline 
Concentric  & Radius of & Radius of & Radius of & Radius of\tabularnewline
circle & the target  & the target  & the matching & the matching\tabularnewline
reproduction & circle 1 & circle 2 & circle 1 & circle 2\tabularnewline
\hline 
Floral shape  & Amplitude 1 & Amplitude 2 & Amplitude 1  & Amplitude 2\tabularnewline
reproduction, & of the & of the & of the  & of the\tabularnewline
see \eqref{eq:-5-1} & target shape & target shape & matching & matching\tabularnewline
 &  &  & shape & shape\tabularnewline
\hline 
\end{tabular}
\end{table}

\end{singlespace}

\subsection{Stimuli and Procedure}

\begin{singlespace}
Visual stimuli consisting of curves and dots were presented on a flat-panel
monitor. The diameter of the dots and the width of the curves was
5 pixels (px). The stimuli were grayish-white on a comfortably low
intensity background. The participants viewed the stimuli in darkness
using a chin rest with a forehead support at the distance of 90 cm
from the monitor, making 1 screen pixel approximately 62 sec arc. 
\end{singlespace}

\subsubsection{Experiment 1}

\begin{singlespace}
In Experiment 1(a), each trial began with presenting two circles with
a dot in the first quadrant of each circle (as shown in Figure 1(a)).
The dot in the upper left circle was fixed at one of randomly chosen
six positions. These six positions contained a $2\times2$ ``rectangular''
sub-design: \{32 px, 64 px\}$\times$\{32 px, 64 px\} and a $2\times2$
``polar'' sub-design: \{53.67 px, 71.55 px\}$\times$\{$63.43$
deg, $26.57$ deg\}. The coordinates were recorded using the center
of the circle as the origin. The position adjustable dot was in the
bottom right circle. The task was to move the bottom right dot by
rotating the trackball to a position that matched that of the fixed
one. There were 1200 trials overall. 
\end{singlespace}

Experiment 1(b) was identical to Experiment 1(a) except the horizontal
coordinate and vertical coordinate of the target dot were random integers
drawn from the interval {[}20 px, 80 px). This ``rectangular'' design
also contained a ``polar'' sub-design {[}40 px, 90 px)$\times${[}$30$
deg, $60$ deg). The overall number of trials for the ``rectangular''
design was 1800, for the polar sub-design about 900. 

\subsubsection{Experiment 2}

\begin{singlespace}
In each trial of Experiment 2(a), the target stimulus on the left
consisted of two concentric circles and a dot in their center. The
radii of circle 1 and circle 2 were randomly chosen from the sets
\{16 px, 56 px, 64 px\} and \{48 px, 72 px, 80 px\}, respectively.
At the beginning of each trial the right stimulus was a dot. The participants
had to reproduce the target stimulus by rotating the trackball to
``blow up'' two circles from that dot one by one. They had the freedom
to produce the inner or the outer circle first. Once the first matching
circle was produced, the participants clicked a button on the trackball
to confirm this circle and then the program enabled them to ``blow
up'' the other circle. After the second circle was created, the trial
was terminated by clicking the same button. There were 1800 trials
overall. 
\end{singlespace}

Experiment 2(b) was identical to Experiment 2(a) except that in each
trial the radii of the target circles were randomly chosen from four
possibilities \{12 px, 24 px\}$\times$\{18 px, 30 px\}. There were
1600 trials overall. 

Experiment 2(c) was identical to Experiment 2(a) except that in each
trial the radii of the target circles were numbers randomly chosen
from {[}18 px, 48 px) $\times${[}56 px, 86 px). There were 1800 trials
overall. 

\subsubsection{Experiment 3}

\begin{singlespace}
Two floral shapes (Figure 1(c)) were presented simultaneously in each
trial in Experiment 3(a). The target one was on the left. The right
one was modifiable. Each floral shape was generated by a function
\begin{align}
x & =\cos(.02\pi\Delta)\text{[}70+\alpha\cos(.06\pi\Delta)+\beta\cos(.1\pi\Delta)],\label{eq:-5-1}\\
y & =\sin(.02\pi\Delta)\text{[}70+\alpha\cos(.06\pi\Delta)+\beta\cos(.1\pi\Delta)],\nonumber 
\end{align}
where $\Delta$ is polar angle and $x$ and $y$ are the horizontal
and vertical coordinates (in pixels). For a matching floral shape,
$\alpha,\beta$ are replaced with $A,B$, respectively. The amplitudes
$\alpha,\beta$ of the target shape were randomly chosen from the
sets \{-18 px, 10 px, 14 px\} and \{-16 px, -12 px, 20 px\}, respectively.
The two amplitudes of the right shape were randomly initialized from
the interval {[}-35 px, 35 px). The participants were asked to match
the left shape by modifying the right shape by rotating the trackball.
There were 1800 trials overall. 
\end{singlespace}

Experiment 3(b) was identical to Experiment 3(a) except that the two
amplitudes of the target shape were randomly chosen numbers from the
interval {[}-30 px, 30 px).
\begin{singlespace}

\section{Results}
\end{singlespace}

\begin{singlespace}
In each experiment the matching points that were too far from the
target values were considered outliers and they were removed from
data analysis. The outliers made less than 1\% of all data. Ref. \cite{DZK2015-1}
briefly reported how contextuality for cyclic systems of rank 4 was
tested using the data collected from our seven experiments. In this
paper, we present the contextuality test for rank 4 in greater details,
and add the analyses for cyclic systems of ranks 6 and 8, using the
same data.
\end{singlespace}

\bigskip{}

\begin{singlespace}

\subsection{Testing Contextuality for Rank 4}
\end{singlespace}

A cyclic system of rank 4 can be represented by four contexts 

\begin{equation}
\begin{array}{cccc}
\mathrm{Context\,}1 & \mathrm{Context\,}2 & \mathrm{Context\,}3 & \mathrm{Context\,}4\\
(\alpha_{1},\beta_{1}) & (\beta_{1},\alpha_{2}) & (\alpha_{2},\beta_{2}) & (\beta_{2},\alpha_{1})
\end{array}.\label{eq: cyclic 4-1}
\end{equation}
To form such a system, we chose $\left\{ \alpha_{1},\alpha_{2}\right\} \times\left\{ \beta_{1},\beta_{2}\right\} =\left\{ 32\,\mathrm{px},64\,\mathrm{px}\right\} \times\left\{ 32\,\mathrm{px},64\,\mathrm{px}\right\} $
for the ``rectangular'' sub-design of Experiment 1(a). The ``polar''
sub-designs of Experiment 1(a) and Experiment 2(b) also have $2\times2$
structures, and they were presented as cyclic systems analogously.
Experiment 2(a) and Experiment 3(a) have $3\times3$ factorial designs.
We extracted 9 cyclic systems of rank 4 from each of them by selecting
two $\alpha$'s and two $\beta$'s from the sets of $\alpha$ and
$\beta$. The ``rectangular'' design of Experiment 1(b) and the
``polar'' sub-designs of Experiment 1(b), Experiment 2(c), and Experiment
3(b) have external factors spanning certain intervals. In order to
have a cyclic system of rank 4, each interval was dichotomized into
two subintervals. For instance, four experimental conditions $(\alpha_{i_{1}},\beta_{i_{2}})$,
$i_{1},i_{2}\in\left\{ 1,2\right\} $, are formed in the ``rectangular''
design of Experiment 1(b) if one chooses $\alpha_{1}=[20\,\mathrm{px},50\,\mathrm{px})$,
$\alpha_{2}=[50\,\mathrm{px},80\,\mathrm{px})$, $\beta_{1}=[20\,\mathrm{px},50\,\mathrm{px})$,
and $\beta_{2}=[50\,\mathrm{px},80\,\mathrm{px})$. Of course other
cut-off points can be chosen to dichotomize the intervals. In this
paper, we only report the results from the midpoint-dichotomized data
sets. 

Irrespective of the experiment, the random outputs $A_{i_{1}i_{2}},B_{i_{1}i_{2}}$
should each be dichotomized. The two values for each random variable
were defined by choosing a value $a_{i_{1}}$ and a value $b_{i_{2}}$
and computing
\begin{equation}
\begin{array}{c}
A_{1i_{2}}^{*}=\left\{ \begin{array}{ccc}
+1 & \mathrm{if} & A_{1i_{2}}>a_{1}\\
-1 & \mathrm{if} & A_{1i_{2}}\leq a_{1}
\end{array}\right.,\quad A_{2i_{2}}^{*}=\left\{ \begin{array}{ccc}
+1 & \mathrm{if} & A_{2i_{2}}>a_{2}\\
-1 & \mathrm{if} & A_{2i_{2}}\leq a_{2}
\end{array}\right.,\\
\\
B_{i_{1}1}^{*}=\left\{ \begin{array}{ccc}
+1 & \mathrm{if} & B_{i_{1}1}>b_{1}\\
-1 & \mathrm{if} & B_{i_{1}1}\leq b_{1}
\end{array}\right.,\quad B_{i_{1}2}^{*}=\left\{ \begin{array}{ccc}
+1 & \mathrm{if} & B_{i_{1}2}>b_{2}\\
-1 & \mathrm{if} & B_{i_{1}2}\leq b_{2}
\end{array}\right..
\end{array}
\end{equation}
We chose a value $a_{1}$ as any integer (in pixels) between max(min$A_{11}$,
min$A_{12}$) and min(max$A_{11}$, max$A_{12}$), $b_{1}$ as any
integer (in pixels or degrees) between max(min$B_{11}$, min$B_{21}$)
and min(max$B_{11}$, max$B_{21}$), and analogously for $a_{2}$
and $b_{2}$. The total number of the rank-4 systems thus formed varied
from 3024 to 11,663,568 per experiment per participant. For each choice
of the quadruple, we applied the test \eqref{delta C}-\eqref{delta C add}
to the distributions of the obtained $A^{*}$ and $B^{*}$ variables.
No positive $\triangle C$ was observed, indicating the absence of
contextuality in the rank 4 cyclic system for each participant in
each experiment.

We present an example to illustrate how the test of (non)contextuality
was conducted. For participant P3 in the ``polar'' sub-design of
Experiment 1(a), one choice of the quadruple was $\left(a_{1},a_{2},b_{1},b_{2}\right)=$
(72 px, 67 px, 60 deg, 23 deg). The distributions of the random outputs
for the four contexts indexed as in \eqref{eq: cyclic 4-1} are presented
in Table \ref{tab: Distributions-of-the}, where the numbers in the
grids are joint probabilities and the numbers outside are marginal
probabilities.

\begin{table}
\begin{centering}
\caption{Distributions of the random outputs for the cyclic system of rank
4, P3 in the ``polar'' sub-design of Experiment 1(a).\label{tab: Distributions-of-the}}
\par\end{centering}
\begin{centering}
\begin{tabular}{|c|c|c|c}
\cline{1-3} 
Context 1 & $B_{11}>b_{1}$ & $B_{11}\leq b_{1}$ & \tabularnewline
\cline{1-3} 
$A_{11}>a_{1}$ & .0056 & 0 & .0056\tabularnewline
\cline{1-3} 
$A_{11}\leq a_{1}$ & .3944 & .6 & .9944\tabularnewline
\cline{1-3} 
\multicolumn{1}{c}{} & \multicolumn{1}{c}{.4} & \multicolumn{1}{c}{.6} & \tabularnewline
\end{tabular}$\quad$%
\begin{tabular}{|c|c|c|c}
\cline{1-3} 
Context 2 & $B_{21}>b_{1}$ & $B_{21}\leq b_{1}$ & \tabularnewline
\cline{1-3} 
$A_{21}>a_{2}$ & .6403 & .3399 & .9802\tabularnewline
\cline{1-3} 
$A_{21}\leq a_{2}$ & .0099 & .0099 & .0198\tabularnewline
\cline{1-3} 
\multicolumn{1}{c}{} & \multicolumn{1}{c}{.6502} & \multicolumn{1}{c}{.3498} & \tabularnewline
\end{tabular}
\par\end{centering}
\centering{}%
\begin{tabular}{|c|c|c|c}
\cline{1-3} 
Context 3 & $B_{22}>b_{2}$ & $B_{22}\leq b_{2}$ & \tabularnewline
\cline{1-3} 
$A_{22}>a_{2}$ & .5789 & .4167 & .9956\tabularnewline
\cline{1-3} 
$A_{22}\leq a_{2}$ & .0044 & 0 & .0044\tabularnewline
\cline{1-3} 
\multicolumn{1}{c}{} & \multicolumn{1}{c}{.5833} & \multicolumn{1}{c}{.4167} & \tabularnewline
\end{tabular}$\quad$%
\begin{tabular}{|c|c|c|c}
\cline{1-3} 
Context 4 & $B_{12}>b_{2}$ & $B_{12}\leq b_{2}$ & \tabularnewline
\cline{1-3} 
$A_{12}>a_{1}$ & .0273 & .0219 & .0492\tabularnewline
\cline{1-3} 
$A_{12}\leq a_{1}$ & .4699 & .4809 & .9508\tabularnewline
\cline{1-3} 
\multicolumn{1}{c}{} & \multicolumn{1}{c}{.4972} & \multicolumn{1}{c}{.5028} & \tabularnewline
\end{tabular}
\end{table}
 We have, in reference to \eqref{delta C}-\eqref{delta C add}
\begin{align*}
\mathrm{s_{1}}\left(\left\langle A_{11}^{*}B_{11}^{*}\right\rangle ,\left\langle B_{21}^{*}A_{21}^{*}\right\rangle ,\left\langle A_{22}^{*}B_{22}^{*}\right\rangle ,\left\langle B_{12}^{*}A_{12}^{*}\right\rangle \right) & =\mathrm{s_{1}}\left(.2112,.3004,.1578,.0164\right)=0.653,
\end{align*}
\begin{align*}
\ICC & =\left|\left\langle A_{11}^{*}\right\rangle -\left\langle A_{12}^{*}\right\rangle \right|+\left|\left\langle B_{11}^{*}\right\rangle -\left\langle B_{21}^{*}\right\rangle \right|+\left|\left\langle A_{21}^{*}\right\rangle -\left\langle A_{22}^{*}\right\rangle \right|+\left|\left\langle B_{22}^{*}\right\rangle -\left\langle B_{12}^{*}\right\rangle \right|\\
 & =\left|(-.9016)-(-.9888)\right|+\left|(-.2)-.3004\right|+\left|.9604-.9912\right|+\left|.1666-(-.0056)\right|\\
 & =.7906.
\end{align*}
With $2n-2=4-2=2$ we obtain
\[
\Delta C=-2.1376<0,
\]
no evidence of contextuality.

\subsection{Testing Contextuality for Rank 6}

Both Experiment 2(a) and Experiment 3(a) have $3\times$3 designs:
$\left\{ \alpha_{1},\alpha_{2},\alpha_{3}\right\} \times\left\{ \beta_{1},\beta_{2},\beta_{3}\right\} $.
From each of them we extracted one cyclic system of rank 6,
\begin{equation}
\begin{array}{cccccc}
\mathrm{Context\,}1 & \mathrm{Context\,}2 & \mathrm{Context\,}3 & \mathrm{Context\,}4 & \textnormal{Context}\,5 & \textnormal{Context}\,6\\
(\alpha_{1},\beta_{1}) & (\beta_{1},\alpha_{2}) & (\alpha_{2},\beta_{2}) & (\beta_{2},\alpha_{3}) & (\alpha_{3},\beta_{3}) & (\beta_{3},\alpha_{1})
\end{array},\label{cyclic 6}
\end{equation}
and labeled the random outputs $A,B$ accordingly. 

The ``rectangular'' design of Experiment 1(b), the ``polar'' sub-designs
of Experiment 1(b), Experiment 2(c), and Experiment 3(b) are the systems
with quasi-continuous factors. These factors were discretized into
three levels by using the one-third quantile and the two-third quantile
of each interval as cut-off points. 

Again, the random outputs should be dichotomized in each experiment.
We chose a value $a_{i_{1}}$ and a value $b_{i_{2}}$, $i_{1},i_{2}\in\left\{ 1,2,3\right\} $,
and defined
\begin{equation}
\begin{array}{c}
A_{1i_{2}}^{*}=\left\{ \begin{array}{ccc}
+1 & \mathrm{if} & A_{1i_{2}}>a_{1}\\
-1 & \mathrm{if} & A_{1i_{2}}\leq a_{1}
\end{array}\right.,\quad A_{2i_{2}}^{*}=\left\{ \begin{array}{ccc}
+1 & \mathrm{if} & A_{2i_{2}}>a_{2}\\
-1 & \mathrm{if} & A_{2i_{2}}\leq a_{2}
\end{array}\right.,\\
\\
A_{3i_{2}}^{*}=\left\{ \begin{array}{ccc}
+1 & \mathrm{if} & A_{3i_{2}}>a_{3}\\
-1 & \mathrm{if} & A_{3i_{2}}\leq a_{3}
\end{array}\right.,\quad B_{i_{1}1}^{*}=\left\{ \begin{array}{ccc}
+1 & \mathrm{if} & B_{i_{1}1}>b_{1}\\
-1 & \mathrm{if} & B_{i_{1}1}\leq b_{1}
\end{array}\right.,\\
\\
B_{i_{1}2}^{*}=\left\{ \begin{array}{ccc}
+1 & \mathrm{if} & B_{i_{1}2}>b_{2}\\
-1 & \mathrm{if} & B_{i_{1}2}\leq b_{2}
\end{array}\right.,\quad B_{i_{1}3}^{*}=\left\{ \begin{array}{ccc}
+1 & \mathrm{if} & B_{i_{1}3}>b_{3}\\
-1 & \mathrm{if} & B_{i_{1}3}\leq b_{3}
\end{array}\right..
\end{array}
\end{equation}

We chose $a_{1}$ as any integer between max(min$A_{11}$, min$A_{13}$)
and min(max$A_{11}$, max$A_{13}$), $b_{1}$ as any integer between
max(min$B_{11}$, min$B_{21}$) and min(max$B_{11}$, max$B_{21}$),
and analogously for $a_{2}$, $a_{3}$, $b_{2}$, and $b_{3}$ for
the experiments with discrete factor points (Experiment 2(a) and Experiment
3(a)). For the experiments with quasi-continuous factors, we chose
$a_{1},a_{2},a_{3},b_{1},b_{2},b_{3}$ as every third integer within
the corresponding range . The total number of the rank-6 systems thus
formed varied from 18,000 to 31,905,600 per experiment per participant.
For each such choice of the sextuple $\left(a_{1},a_{2},a_{3},b_{1},b_{2},b_{3}\right)$
we conducted the test \eqref{delta C}-\eqref{delta C add}. No positive
$\triangle C$ was observed for the systems of rank 6 we investigated.

\begin{table}
\begin{centering}
\caption{Distributions of the random outputs for the cyclic system of rank
6, P1 in Experiment 2(a).\label{tab: Distributions-of-the-1}}
\par\end{centering}
\begin{centering}
\begin{tabular}{|c|c|c|c}
\cline{1-3} 
Context 1 & $B_{11}>b_{1}$ & $B_{11}\leq b_{1}$ & \tabularnewline
\cline{1-3} 
$A_{11}>a_{1}$ & .2124 & .2487 & .4611\tabularnewline
\cline{1-3} 
$A_{11}\leq a_{1}$ & .1917 & .3472 & .5389\tabularnewline
\cline{1-3} 
\multicolumn{1}{c}{} & \multicolumn{1}{c}{.4041} & \multicolumn{1}{c}{.5959} & \tabularnewline
\end{tabular}$\quad$%
\begin{tabular}{|c|c|c|c}
\cline{1-3} 
Context 2 & $B_{21}>b_{1}$ & $B_{21}\leq b_{1}$ & \tabularnewline
\cline{1-3} 
$A_{21}>a_{2}$ & .2353 & .1041 & .3394\tabularnewline
\cline{1-3} 
$A_{21}\leq a_{2}$ & .1538 & .5068 & .6606\tabularnewline
\cline{1-3} 
\multicolumn{1}{c}{} & \multicolumn{1}{c}{.3891} & \multicolumn{1}{c}{.6109} & \tabularnewline
\end{tabular}
\par\end{centering}
\begin{centering}
\begin{tabular}{|c|c|c|c}
\cline{1-3} 
Context 3 & $B_{22}>b_{2}$ & $B_{22}\leq b_{2}$ & \tabularnewline
\cline{1-3} 
$A_{22}>a_{2}$ & .1221 & .0814 & .2035\tabularnewline
\cline{1-3} 
$A_{22}\leq a_{2}$ & .1628 & .6337 & .7965\tabularnewline
\cline{1-3} 
\multicolumn{1}{c}{} & \multicolumn{1}{c}{.2849} & \multicolumn{1}{c}{.7151} & \tabularnewline
\end{tabular}$\quad$%
\begin{tabular}{|c|c|c|c}
\cline{1-3} 
Context 4 & $B_{32}>b_{2}$ & $B_{32}\leq b_{2}$ & \tabularnewline
\cline{1-3} 
$A_{32}>a_{3}$ & .2703 & .0586 & .3288\tabularnewline
\cline{1-3} 
$A_{32}\leq a_{3}$ & .1982 & .4730 & .6712\tabularnewline
\cline{1-3} 
\multicolumn{1}{c}{} & \multicolumn{1}{c}{.4685} & \multicolumn{1}{c}{.5316} & \tabularnewline
\end{tabular}
\par\end{centering}
\centering{}%
\begin{tabular}{|c|c|c|c}
\cline{1-3} 
Context 5 & $B_{33}>b_{3}$ & $B_{33}\leq b_{3}$ & \tabularnewline
\cline{1-3} 
$A_{33}>a_{3}$ & .0702 & .0468 & .1170\tabularnewline
\cline{1-3} 
$A_{33}\leq a_{3}$ & .0409 & .8421 & .8830\tabularnewline
\cline{1-3} 
\multicolumn{1}{c}{} & \multicolumn{1}{c}{.1111} & \multicolumn{1}{c}{.8889} & \tabularnewline
\end{tabular}$\quad$%
\begin{tabular}{|c|c|c|c}
\cline{1-3} 
Context 6 & $B_{13}>b_{3}$ & $B_{13}\leq b_{3}$ & \tabularnewline
\cline{1-3} 
$A_{13}>a_{1}$ & .1321 & .1981 & .3302\tabularnewline
\cline{1-3} 
$A_{13}\leq a_{1}$ & .1651 & .5047 & .6698\tabularnewline
\cline{1-3} 
\multicolumn{1}{c}{} & \multicolumn{1}{c}{.2972} & \multicolumn{1}{c}{.7028} & \tabularnewline
\end{tabular}
\end{table}

We present an example of how the test \eqref{delta C}-\eqref{delta C add}
was conducted. For participant P1 in Experiment 2(a), in which $\left\{ \alpha_{1},\alpha_{2},\alpha_{3}\right\} \times\left\{ \beta_{1},\beta_{2},\beta_{3}\right\} =$
\{16 px, 56 px, 64 px\} $\times$ \{48 px, 72 px, 80 px\}, one choice
of the sextuple was $\left(a_{1},a_{2},a_{3},b_{1},b_{2},b_{3}\right)=$
(16 px, 56 px, 64 px, 48 px, 72 px, 80 px). The distributions of the
random outputs for the six contexts indexed as in \eqref{cyclic 6}
are presented in Table \ref{tab: Distributions-of-the-1}.

We have
\begin{align*}
\begin{array}{c}
\mathrm{s_{1}}\left(\left\langle A_{11}^{*}B_{11}^{*}\right\rangle ,\left\langle B_{21}^{*}A_{21}^{*}\right\rangle ,\left\langle A_{22}^{*}B_{22}^{*}\right\rangle ,\left\langle B_{32}^{*}A_{32}^{*}\right\rangle ,\left\langle A_{33}^{*}B_{33}^{*}\right\rangle ,\left\langle B_{13}^{*}A_{13}^{*}\right\rangle \right)\\
=\mathrm{s_{1}}\left(.1192,.4843,.5116,.4865,.8246,.2736\right)=2.4613,
\end{array}
\end{align*}
\begin{align*}
\ICC= & \left|\left\langle A_{13}^{*}\right\rangle -\left\langle A_{11}^{*}\right\rangle \right|+\left|\left\langle B_{11}^{*}\right\rangle -\left\langle B_{21}^{*}\right\rangle \right|+\left|\left\langle A_{22}^{*}\right\rangle -\left\langle A_{21}^{*}\right\rangle \right|+\left|\left\langle B_{22}^{*}\right\rangle -\left\langle B_{32}^{*}\right\rangle \right|\\
 & +\left|\left\langle A_{32}^{*}\right\rangle -\left\langle A_{33}^{*}\right\rangle \right|+\left|\left\langle B_{33}^{*}\right\rangle -\left\langle B_{13}^{*}\right\rangle \right|\\
= & \left|-.0778-(-.3396)\right|+\left|-.1918-(-.2218)\right|+\left|-.3212-(-.593)\right|\\
 & +\left|-.4302-(-.0632)\right|+\left|-.3424-(-.7660)\right|+\left|-.7778-(-.4056)\right|\\
= & .2618+0.03+.2718+.367+.4236+.3722=1.7264,
\end{align*}
whence
\[
\Delta C=2.4613-1.7264-(6-4)=-3.2651<0,
\]
no evidence of contextuality.

\subsection{Testing Contextuality for Rank 8}

The ``rectangular'' design of Experiment 1(b), the ``polar'' sub-designs
of Experiment 1(b), Experiment 2(c), and Experiment 3(b) have quasi-continuous
factors. Each factor in each experiment was discretized into four
levels in order to form a rank 8 cyclic system. Three points should
be chosen for each factor to make this discretization. We chose the
first quartile point, the second quartile (median) point, and the
third quartile point of each interval. A cyclic system of rank 8 was
extracted from each experiment: 
\begin{equation}
\begin{array}{cccccccc}
\mathrm{Context\,}1 & \mathrm{Context\,}2 & \mathrm{Context\,}3 & \mathrm{Context\,}4 & \mathrm{Context\,5} & \mathrm{Context\,6} & \mathrm{Context}\,7 & \mathrm{Context}\,8\\
(\alpha_{1},\beta_{1}) & (\beta_{1},\alpha_{2}) & (\alpha_{2},\beta_{2}) & (\beta_{2},\alpha_{3}) & (\alpha_{3},\beta_{3}) & (\beta_{3},\alpha_{4}) & (\alpha_{4},\beta_{4}) & (\beta_{4},\alpha_{1})
\end{array},\label{cyclic 8}
\end{equation}
with the random outputs labeled accordingly. 

To dichotomize the outputs, we chose a value $a_{i_{1}}$ and a value
$b_{i_{2}}$, $i_{1},i_{2}\in\left\{ 1,2,3,4\right\} $ to define
\begin{equation}
\begin{array}{c}
A_{1i_{2}}^{*}=\left\{ \begin{array}{ccc}
+1 & \mathrm{if} & A_{1i_{2}}>a_{1}\\
-1 & \mathrm{if} & A_{1i_{2}}\leq a_{1}
\end{array}\right.,\quad A_{2i_{2}}^{*}=\left\{ \begin{array}{ccc}
+1 & \mathrm{if} & A_{2i_{2}}>a_{2}\\
-1 & \mathrm{if} & A_{2i_{2}}\leq a_{2}
\end{array}\right.,\\
\\
A_{3i_{2}}^{*}=\left\{ \begin{array}{ccc}
+1 & \mathrm{if} & A_{3i_{2}}>a_{3}\\
-1 & \mathrm{if} & A_{3i_{2}}\leq a_{3}
\end{array}\right.,\quad A_{4i_{2}}^{*}=\left\{ \begin{array}{ccc}
+1 & \mathrm{if} & A_{4i_{2}}>a_{4}\\
-1 & \mathrm{if} & A_{4i_{2}}\leq a_{4}
\end{array}\right.,\\
\\
B_{i_{1}1}^{*}=\left\{ \begin{array}{ccc}
+1 & \mathrm{if} & B_{i_{1}1}>b_{1}\\
-1 & \mathrm{if} & B_{i_{1}1}\leq b_{1}
\end{array}\right.,\quad B_{i_{1}2}^{*}=\left\{ \begin{array}{ccc}
+1 & \mathrm{if} & B_{i_{1}2}>b_{2}\\
-1 & \mathrm{if} & B_{i_{1}2}\leq b_{2}
\end{array}\right.,\\
\\
B_{i_{1}3}^{*}=\left\{ \begin{array}{ccc}
+1 & \mathrm{if} & B_{i_{1}3}>b_{3}\\
-1 & \mathrm{if} & B_{i_{1}3}\leq b_{3}
\end{array}\right.,\quad B_{i_{1}4}^{*}=\left\{ \begin{array}{ccc}
+1 & \mathrm{if} & B_{i_{1}4}>b_{4}\\
-1 & \mathrm{if} & B_{i_{1}4}\leq b_{4}
\end{array}\right..
\end{array}
\end{equation}

For each rank 8 cyclic system, we chose $a_{1}$ as every sixth integer
between max(min$A_{11}$, min$A_{14}$) and min(max$A_{11}$, max$A_{14}$),
$b_{1}$ as every sixth integer between max(min$B_{11}$, min$B_{21}$)
and min(max$B_{11}$, max$B_{21}$), and analogously for $a_{2}$,
$a_{3}$, $a_{4}$, $b_{2}$, $b_{3}$, and $b_{4}$. The total number
of the rank-8 systems thus formed varied from 432 to 6,453,888 per
experiment per participant. For each thus obtained octuple we conducted
the test \eqref{delta C}-\eqref{delta C add}. No positive $\triangle C$
was observed in all the investigated cyclic systems of rank 8. 

To give an example, for participant P4 in Experiment 3(b), one choice
of the octuple was $\left(a_{1},a_{2},a_{3},a_{4},b_{1},b_{2},b_{3},b_{4}\right)=$
(-21 px, -6 px, 6 px, 21 px, -21 px, -9 px, 9 px, 21 px). The distributions
of the random outputs for the eight contexts indexed as in \eqref{cyclic 8}
are presented in Table \ref{tab: Distributions-of-the-2}.

\begin{table}
\begin{centering}
\caption{Distributions of the random outputs for the cyclic system of rank
8, P4 in Experiment 3(b).\label{tab: Distributions-of-the-2}}
\par\end{centering}
\begin{centering}
\begin{tabular}{|c|c|c|c}
\cline{1-3} 
Context 1 & $B_{11}>b_{1}$ & $B_{11}\leq b_{1}$ & \tabularnewline
\cline{1-3} 
$A_{11}>a_{1}$ & .1532 & .2823 & .4355\tabularnewline
\cline{1-3} 
$A_{11}\leq a_{1}$ & .1855 & .3790 & .5645\tabularnewline
\cline{1-3} 
\multicolumn{1}{c}{} & \multicolumn{1}{c}{.3387} & \multicolumn{1}{c}{.6613} & \tabularnewline
\end{tabular}$\quad$%
\begin{tabular}{|c|c|c|c}
\cline{1-3} 
Context 2 & $B_{21}>b_{1}$ & $B_{21}\leq b_{1}$ & \tabularnewline
\cline{1-3} 
$A_{21}>a_{2}$ & .1619 & .2667 & .4286\tabularnewline
\cline{1-3} 
$A_{21}\leq a_{2}$ & .1905 & .3810 & .5715\tabularnewline
\cline{1-3} 
\multicolumn{1}{c}{} & \multicolumn{1}{c}{.3524} & \multicolumn{1}{c}{.6477} & \tabularnewline
\end{tabular}
\par\end{centering}
\begin{centering}
\begin{tabular}{|c|c|c|c}
\cline{1-3} 
Context 3 & $B_{22}>b_{2}$ & $B_{22}\leq b_{2}$ & \tabularnewline
\cline{1-3} 
$A_{22}>a_{2}$ & .2759 & .2155 & .4914\tabularnewline
\cline{1-3} 
$A_{22}\leq a_{2}$ & .2586 & .2500 & .5086\tabularnewline
\cline{1-3} 
\multicolumn{1}{c}{} & \multicolumn{1}{c}{.5345} & \multicolumn{1}{c}{.4655} & \tabularnewline
\end{tabular}$\quad$%
\begin{tabular}{|c|c|c|c}
\cline{1-3} 
Context 4 & $B_{32}>b_{2}$ & $B_{32}\leq b_{2}$ & \tabularnewline
\cline{1-3} 
$A_{32}>a_{3}$ & .4130 & .1739 & .5869\tabularnewline
\cline{1-3} 
$A_{32}\leq a_{3}$ & .1957 & .2174 & .4131\tabularnewline
\cline{1-3} 
\multicolumn{1}{c}{} & \multicolumn{1}{c}{.6087} & \multicolumn{1}{c}{.3913} & \tabularnewline
\end{tabular}
\par\end{centering}
\begin{centering}
\begin{tabular}{|c|c|c|c}
\cline{1-3} 
Context 5 & $B_{33}>b_{3}$ & $B_{33}\leq b_{3}$ & \tabularnewline
\cline{1-3} 
$A_{33}>a_{3}$ & .2736 & .3208 & .5944\tabularnewline
\cline{1-3} 
$A_{33}\leq a_{3}$ & .1604 & .2453 & .4057\tabularnewline
\cline{1-3} 
\multicolumn{1}{c}{} & \multicolumn{1}{c}{.4340} & \multicolumn{1}{c}{.5661} & \tabularnewline
\end{tabular}$\quad$%
\begin{tabular}{|c|c|c|c}
\cline{1-3} 
Context 6 & $B_{43}>b_{3}$ & $B_{43}\leq b_{3}$ & \tabularnewline
\cline{1-3} 
$A_{43}>a_{4}$ & .2460 & .3095 & .5555\tabularnewline
\cline{1-3} 
$A_{43}\leq a_{4}$ & .1667 & .2778 & .4445\tabularnewline
\cline{1-3} 
\multicolumn{1}{c}{} & \multicolumn{1}{c}{.4127} & \multicolumn{1}{c}{.5873} & \tabularnewline
\end{tabular}
\par\end{centering}
\centering{}%
\begin{tabular}{|c|c|c|c}
\cline{1-3} 
Context 7 & $B_{44}>b_{4}$ & $B_{44}\leq b_{4}$ & \tabularnewline
\cline{1-3} 
$A_{44}>a_{4}$ & .3209 & .3134 & .6343\tabularnewline
\cline{1-3} 
$A_{44}\leq a_{4}$ & .1493 & .2164 & .3657\tabularnewline
\cline{1-3} 
\multicolumn{1}{c}{} & \multicolumn{1}{c}{.4702} & \multicolumn{1}{c}{.5298} & \tabularnewline
\end{tabular}$\quad$%
\begin{tabular}{|c|c|c|c}
\cline{1-3} 
Context 8 & $B_{14}>b_{4}$ & $B_{14}\leq b_{4}$ & \tabularnewline
\cline{1-3} 
$A_{14}>a_{1}$ & .1619 & .2571 & .4190\tabularnewline
\cline{1-3} 
$A_{14}\leq a_{1}$ & .2381 & .3429 & .5810\tabularnewline
\cline{1-3} 
\multicolumn{1}{c}{} & \multicolumn{1}{c}{.4} & \multicolumn{1}{c}{.6} & \tabularnewline
\end{tabular}
\end{table}

We have then
\[
\begin{array}{l}
\mathrm{s_{1}}\left(\left\langle A_{11}^{*}B_{11}^{*}\right\rangle ,\left\langle B_{21}^{*}A_{21}^{*}\right\rangle ,\left\langle A_{22}^{*}B_{22}^{*}\right\rangle ,\left\langle B_{32}^{*}A_{32}^{*}\right\rangle ,\left\langle A_{33}^{*}B_{33}^{*}\right\rangle ,\left\langle B_{43}^{*}A_{43}^{*}\right\rangle ,\left\langle A_{44}^{*}B_{44}^{*}\right\rangle ,\left\langle B_{14}^{*}A_{14}^{*}\right\rangle \right)\\
\\
=\mathrm{s_{1}}\left(.0644\text{,}.0857,.0518,.2608,.0377,.0476,.0746,.0096\right)=.613,
\end{array}
\]
\begin{align*}
\ICC= & \left|\left\langle A_{11}^{*}\right\rangle -\left\langle A_{14}^{*}\right\rangle \right|+\left|\left\langle B_{11}^{*}\right\rangle -\left\langle B_{21}^{*}\right\rangle \right|+\left|\left\langle A_{22}^{*}\right\rangle -\left\langle A_{21}^{*}\right\rangle \right|+\left|\left\langle B_{22}^{*}\right\rangle -\left\langle B_{32}^{*}\right\rangle \right|\\
 & +\left|\left\langle A_{33}^{*}\right\rangle -\left\langle A_{32}^{*}\right\rangle \right|+\left|\left\langle B_{43}^{*}\right\rangle -\left\langle B_{33}^{*}\right\rangle \right|+\left|\left\langle A_{44}^{*}\right\rangle -\left\langle A_{43}^{*}\right\rangle \right|+\left|\left\langle B_{14}^{*}\right\rangle -\left\langle B_{44}^{*}\right\rangle \right|\\
= & \left|-.129-(-.162)\right|+\left|-.3226-(-.2952)\right|+\left|-.0172-(-.1428)\right|+\left|.069-.2174\right|\\
 & +\left|.1888-.1738\right|+\left|-.1746-(-.132)\right|+\left|.2686-(.1110)\right|+\left|-.2-(-.0596)\right|\\
= & .033+.0274+.1256+.1484+.015+.0426+.1576+.1404=.6902,
\end{align*}
whence
\[
\Delta C=.613-.6902-(8-2)=-6.0772<0,
\]
no evidence of contextuality.

\section{Conclusions}

Contextuality-by-default is a mathematical framework that allows to
classify systems as contextual or noncontextual. Experimental data
suggest that the noncontextuality boundaries are generally breached
in quantum physics \cite{KDL2015}. In Refs. \cite{DKCZJ2016,DZK2015-1}
we reviewed several behavioral and social scenarios to conclude that
none of them provided evidence for contextuality. By examining the
psychophysical data collected in our laboratory, we found no contextuality
for cyclic systems of different ranks, including high ranks (6 and
8) that have never been analyzed before. We suspect that it may be
generally true that human and social behaviors are not contextual
in the same sense in which quantum systems are.
\begin{singlespace}

\subsection*{Acknowledgments}
\end{singlespace}

This research has been supported by NSF grant SES-1155956, AFOSR grant
FA9550-14-1-0318.

\end{document}